\documentclass[preprint,amsmath,amssymb,prl,epsf,rotate,doublespace]{revtex4}

\usepackage{graphicx}
\usepackage{dcolumn}
\usepackage{bm}
\usepackage{float}

\begin{document}

\title{Scaling Phenomena in the Growth Dynamics of Scientific Output }


\author{Kaushik Matia$^1$, Luis A. Nunes Amaral$^{1,2}$, Marc Luwel$^3$,
  Henk.~F.~Moed$^4$, H. Eugene~Stanley$^1$}
\affiliation{%
$^1$Center for Polymer Studies and Department of Physics,
  Boston University, Boston, MA 02215\\
$^2$Department of Chemical Engineering,
    Northwestern University, Evanston, IL60208\\
$^3$Ministry of the Flemish Community, Brussels, Belgium\\
$^4$Center for Science and Technology Studies, Leiden Univ, POB 9555,
    NL-2300 RB Leiden, Netherlands\\
}%



\begin{abstract}

We analyze a set of three databases at different levels of aggregation
(i) a database of approximately $10^6$ publications of 247 countries in
the period between 1980--2001. (ii) A database of 508 academic
institutions from European Union (EU) and 408 institutes from USA in the
11 year period between during 1991--2001. (iii) A database comprising of
2330 Flemish authors in the period 1980--2000. At all levels of
aggregation we find that the mean annual growth rates of publications is
independent of the number of publications of the various units
involved. We also find that the standard deviation of the distribution
of annual growth rates decays with the number of publications as a power
law with exponent $\approx 0.3$. These findings are consistent with
those of recent studies of systems such as the size of R\&D funding
budgets of countries, the research publication volumes of US
universities, and the size of business firms.

\end{abstract}

\date{working paper last revised: \today}

\pacs{PACS numbers: 89.90.+n, 05.45.Tp, 05.40.Fb}
\maketitle


\section{Introduction}

One outcome of World War II and the role science and technology (S\&T)
played in that conflict was a heightened awareness on the part of policy
makers of how developments in science and technology affect the
security, economic development, and public good of a nation (Durlauf,
1996; Chandler, 1962; Gort, 1962). Since that time, science and
technology studies focusing on the complex relationships influencing
research, development, and innovation have produced many policy-relevant
results. Vannevar Bush's ground-breaking {\it Report to the President on
a Program for Postwar Scientific Research\/} (based on the linear model
presented in (National Science Board, 2000); Mansfield 1991; Jaffe 1996)
dominated policy thinking in the period after World War II but, within
the knowledge industry, emerging new concepts---such as the national
innovation system---have highlighted the complex interactions between
research, development, and innovation and have clarified their economic
and social relevance (Durlauf, 1995).

It is now clear that R\&D spending decisions e.g., how to partition
funds among disciplines (e.g. weighted toward life sciences or natural
sciences ) or how much to spend on individual projects (e.g.spending for
the human genome project or global warming or renewable sources of
energy) can dramatically impact the pattern of development, strongly
influence which advances occur first and, if strategic decisions are
haphazard, seriously jeopardize the competitiveness of the entire S\&T
system (Pakes, 1996).
These concerns are even more pressing now than they were 50 years ago
because 

\begin{itemize}

\item[{(i)}] the scale of the S\&T systems and the available resources
are now much larger, 

\item[{(ii)}] scientific advances now take place much more rapidly, 

\item[{(iii)}] cutting-edge research today is often multidisciplinary
(e.g., in the new field of bio-informatics, biologists, mathematicians,
and physicists sometimes cooperate and sometimes compete), and

\item[{(iv)}] research results and technological innovations have a
stronger impact on economic growth and competitiveness.

\end{itemize}

To make informed choices, decision makers need information that is
timely, reliable, and clear (Luwel, 1999). To answer these needs, the
field of quantitative S\&T studies has gone through a revolutionary
period (CWTS, 2000) during which many new indicators have been
identified (Garfield, 1979), but in spite of important advances, this is
still an extremely complex project with many unsolved
questions. Indicators are, by definition, retrospective and heuristic
(National Science Board, 2000), and there are many difficulties
associated with the development of indicators (Moed, 1995; Plerou, 1999)
that are general, robust, and applicable (i) across different S\&T
fields, (ii) for different aggregation levels (from research groups to
entire countries), (iii) equally well for input and output measures.

Most bibliometric indicators are one-dimensional i.e., they analyze only
one variable such as R\&D spending, number of publications, number of
citations, or time evolution. Indicators based on these variables (e.g.,
OECD S\&T-indicators, NSF Science and Engineering Indicators, EU Science
and Technology Indicators) are well-known to policy-makers, but to
better understand the underlying processes driving the R\&D system and
how they impact economic development, we need to better understand the
relationships among these variables and thus far, little work has been
done in this area. Appropriate research could produce more complex
indicators that may enable us to more accurately predict the output and
impact of policy changes. Indeed, OECD has already stated that such
``blue sky'' indicators are indispensable policy tools in a knowledge
economy driven by research and technological innovation. The approach
adopted in this paper is inspired by (Derek de Solla Price, 1963) who
conceived science as a {\it physical} system. He aimed at simple laws,
similar to those in planetary physics discovered by Newton. Rather than
applying laws from classical physics, our goal is to develop more
sophisticated R\&D indicators by using concepts and tools recently
developed in the field of statistical physics.  Specifically, we will
apply two of that field's fundamental concepts: scaling and universality
(Stanley, 1999).

\subsection{Scaling and Universality}

The utility of the ``universality'' concept can be explained through an
analogy with the Mendeleev periodic table of atomic elements. During the
last century, Mendeleev noticed that some elements shared similar
physical and chemical properties.  That observation prompted him to
organize the atomic elements known at that time into a table in which
atomic elements with similar properties occupy the same column.  By
organizing the elements into this table, Mendeleev found that some cells
of this periodic table were left empty.  Later, those empty cells were
found to correspond to newly-discovered atomic elements whose chemical
and physical properties were well-predicted by their position in the
table.

Analogously, the study of critical phenomena in statistical physics
has shown that the phase transition of very different systems---e.g., water
at the critical point, a polymer at its collapsing temperature, or a
magnet undergoing a temperature change---could be classified into a few
classes, each class being described by the same scaling functions and
the same scaling laws.

This result motivates a question of fundamental importance: {\it ``Which
features of this microscopic inter-particle force are important for
determining critical-point exponents and scaling functions, and which
are unimportant?"}  This question has been answered for physical
systems, but is still lacking an answer for other systems.  The
discovery of universality in physical systems is also of great practical
interest.  Specifically, when studying a given problem, one may pick the
most tractable system to study and the results one obtains will hold for
all other systems in the same universality class.

Here we extend a recent study by (Plerou 1999; Moed 1999;) and
investigate to what extent the concept of scaling can (i) be used to
study R\&D systems by analyzing the publication output of academic
research institutions and authors and (ii) lead to new and more
sophisticated indicators.  Contrary to technological innovation,
scientific knowledge is a public good and researchers establish
intellectual property for their results by publishing them. The
processes leading to new scientific knowledge are complex and, to a
large extent, driven by a government's R\&D-policy. This policy varies
considerably over countries in areas such as the total public investment
in R\&D, the priority setting between scientific disciplines, the
institutional organization (universities, public research institutes,
etc.) and the way research itself is funded (more or less competitively
driven).

\subsection{Growth of Organizations}

Consider the annual growth rate of an organization's size
\begin{equation}
g(t) \equiv \log \left(\frac{S(t+1)}{S(t)} \right) = \log S(t+1) - \log S(t),
\label{e.grate}
\end{equation}
where $S(t)$ and $S(t+1)$ are the size of the organization being
considered in the years $t$ and $t+1$, respectively. The organization
can be a business firm (Stanley, 1996; Amaral, 1997; Buldyrev, 1997;
Takayasu, 1998; Sutton, 2000; Wyart, 2002), a country (Canning, 1998), a
university research budget (Plerou, 1999), a voluntary social
organization, or a bird species (Keitt, 1998; Keitt, 2002). We expect
that the statistical properties of the growth rate $g$ depend on $S$,
since it is natural that the magnitude of the fluctuations $g$ will
decrease with $S$. We partition the growth rates into groups according
to their sizes to test whether the probability density {\it
conditioned\/} on the size $p(g | S)$ has the same functional form for
all the different size groups (Stanley, 1996; Amaral, 1997; Buldyrev,
1997).

If the conditional distribution of growth rates has a functional form
dependent on S, we expect the standard deviation $\sigma(S)$---which is
a measure of the width of $p(g|S)$---to be dependent on S. Thus, if when we
plot the scaled quantities
\begin{equation}
\sigma(S) p(g/\sigma(S) | S) \qquad \mbox{versus} \qquad g/\sigma(S)
\end{equation}
all $\sigma$ curves from the different size groups collapse onto a
single curve, then $p(g | S)$ follows a universal scaling (Amaral, 1997,
Buldyrev, 1997)
\begin{equation}
p(g | S) \sim \frac{1}{\sigma(S)} ~~ f \left( \frac{g}{\sigma(S)} \right)\,.
\label{e.dist-grate2}
\end{equation}
where $f$ is a symmetric function independent of $S$ of a specific
``tent-shaped'' form. Models (Amaral, 1998; Matia, 2004) discusses how
the tent-shaped form of $f$ can be interpreted by a convolution of a log
normal distributions and a Gaussian distribution. Interestingly, our
studies reveal that $\sigma(S)$ decays as a power law Stanley (1996),
Buldyrev (1997)
\begin{equation}
\sigma(S) \sim S^{-\beta},
\label{e.sigma}
\end{equation}
where $\beta$ is known as the {\it scaling exponent}.


\section{Data For Different Levels of Aggregation}

\subsection{Data of Publication of Countries}

We analyze a database consisting of the total annual publications of 247
countries between 1980--2001. We extract the data from the
CD-ROM version of the Science Citation Index (SCI) published by the
Institute for Scientific Information (ISI) at Philadelphia, USA, founded
by Eugene Garfield.

We count country publications in three distinct ways, which we
illustrate with an example: Consider one publication co-authored by
researchers affiliated with four different institutions in three
different countries. Two of the study's authors are affiliated with a
particular US institution, a third author to a second US institution, a
fourth with a Dutch institution and the last author with a Belgian
institution. For this case, one can define at least four different
assignments of the publication to the three countries involved.

In an ideal case, one would assign fractions of a paper to a country on
the basis of the proportion of authors from each country.  Thus, in the
example, 0.6 publications would be assigned to the US, 0.2 to the
Netherlands and 0.2 to Belgium. However, in the database analyzed,
authors are not tagged to institutions.  Therefore, for multi-authored
papers from different institutions, the distribution of authors among
institutions or countries cannot be determined. In our study,
publications were assigned to countries on the basis of the geographic
location of the authors' institutions rather than that of the authors
themselves.  Thus, three counting schemes can be applied. The first is
denoted as ``fractional count.'' Since two institutions are located in
the US, one in the Netherlands and one in Belgium, 1/2 of the paper is
assigned to US, and 1/4 to each of the other two countries. This count
will be denoted as fractional count throughout this paper. A second,
denoted as ``integer count, type I'' assigns two publications to the US,
one publication to the Netherlands and one publication to Belgium.
Finally, the third, denoted as ``integer count type II'' assigns one
publication each to the US, the Netherlands, and Belgium.

The fractional count definition has the advantage that it conserves the
total number of publications regardless of the number of authors. Our
fractional count is not a perfect solution to the assignment of
publications to countries as it is based on contributing institutions
rather then on individual contributors, but it is the best we are able
to generate with the data available. Moreover, at the level of
countries, differences between a fractional assignment based on
institutions and that based on authors can be expected to level out to a
considerable extent.  The two integer count definitions are important
because they provide a way to determine the weight of national and
international collaborations on the research of a country.  In fact,
type II integer counts reflect international collaboration, and type I
integer counts reflect institutional collaboration both at the national
and the international level.

By considering the three distinct counting methods for publications, we
generate three databases for analysis. From each of these databases we
select the subset of countries which had non-zero publications during
the entire 22-year period. This procedure eliminates 123
countries---some of which were created during the observations period
(due mainly to changes in Eastern Europe and the disintegration of the
USSR) and some with very low publication rates---yielding 124 countries.

\subsection{Data of Publication of Institutes}

We analyze a database consisting of the total annual publications of 508
institutes from European Union (EU) and 408 academic institutions from
USA in the 11 year period between during 1991--2001. Publication by
institutes is recorded according to the fractional counting scheme
described before. Publications were assigned to institutions on the
basis of the institutional affiliations of publishing authors, taking
into account variations in the institutions' name. 

\subsection{Data of Publication of Flemish Authors}

We analyze a database consisting of the total annual publications of
2330 authors between 1980--2001. The database contains articles,
letters, notes and reviews in CDROM version of SCI 1980-2000 Flemish
researchers active in natural and life sciences who during 1991-2000 were
member of a Committee or who submitted a proposal to the Flemish
Research Council FWO-Vlaanderen.

Publication by Flemish authors is recorded in two distinct ways, which
we illustrate with an example: Consider one publication co-authored by
two different researchers. Two different counting schemes can be
applied. The first is denoted as ``fractional count'' where each author
receives a score of $1/2$. A second, denoted as ``integer count''
assigns to each author each author a score of $1$.

\section{Analysis}

\subsection{Countries}

\begin{figure}
\narrowtext
\vspace*{0.1cm}
\begin{center}
\includegraphics[width=0.60\textwidth,angle=-90]{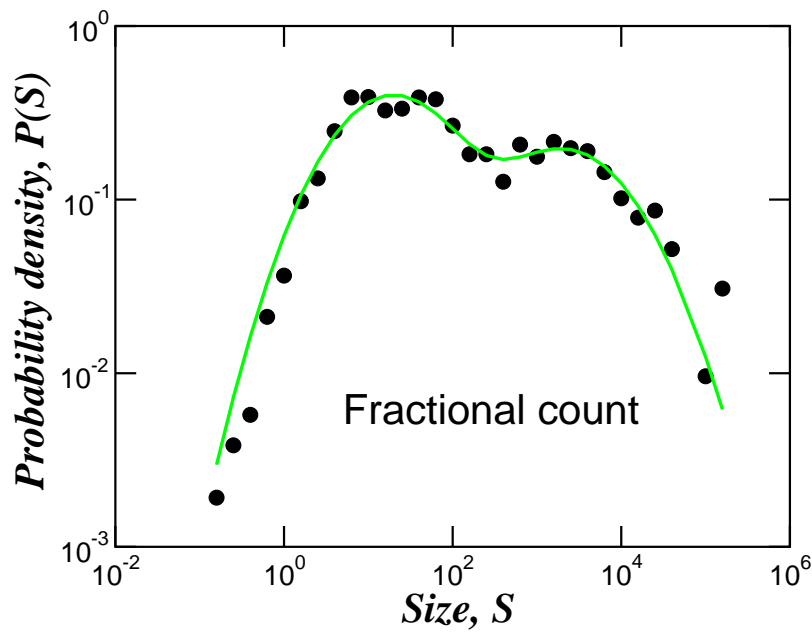}
\end{center}
\vspace*{0.6cm}
\caption{ Histogram of the logarithm of number of publications of 124
countries for the 21-year period between 1980--2001 according to
fractional counting scheme.  The solid line is a Gaussian fit to the
data, which is a prediction of Gibrat's theory. We observe a bi-modal
distribution in the sizes of publication for all different counting
method of countries, which is indicative of two different sectors with
respect to their size. Each of the two sectors grow in a multiplicative
process resulting in a log-normal distribution of sizes. This feature of
size distribution is not observed in the GDP of
countries (Canning, 1998). The two integer counting
scheme also gives similar results.}
\label{Fig2}
\end{figure}

\begin{figure}
\narrowtext
\vspace*{0.6cm}
\begin{center}
\includegraphics[width=0.53\textwidth,angle=-90]{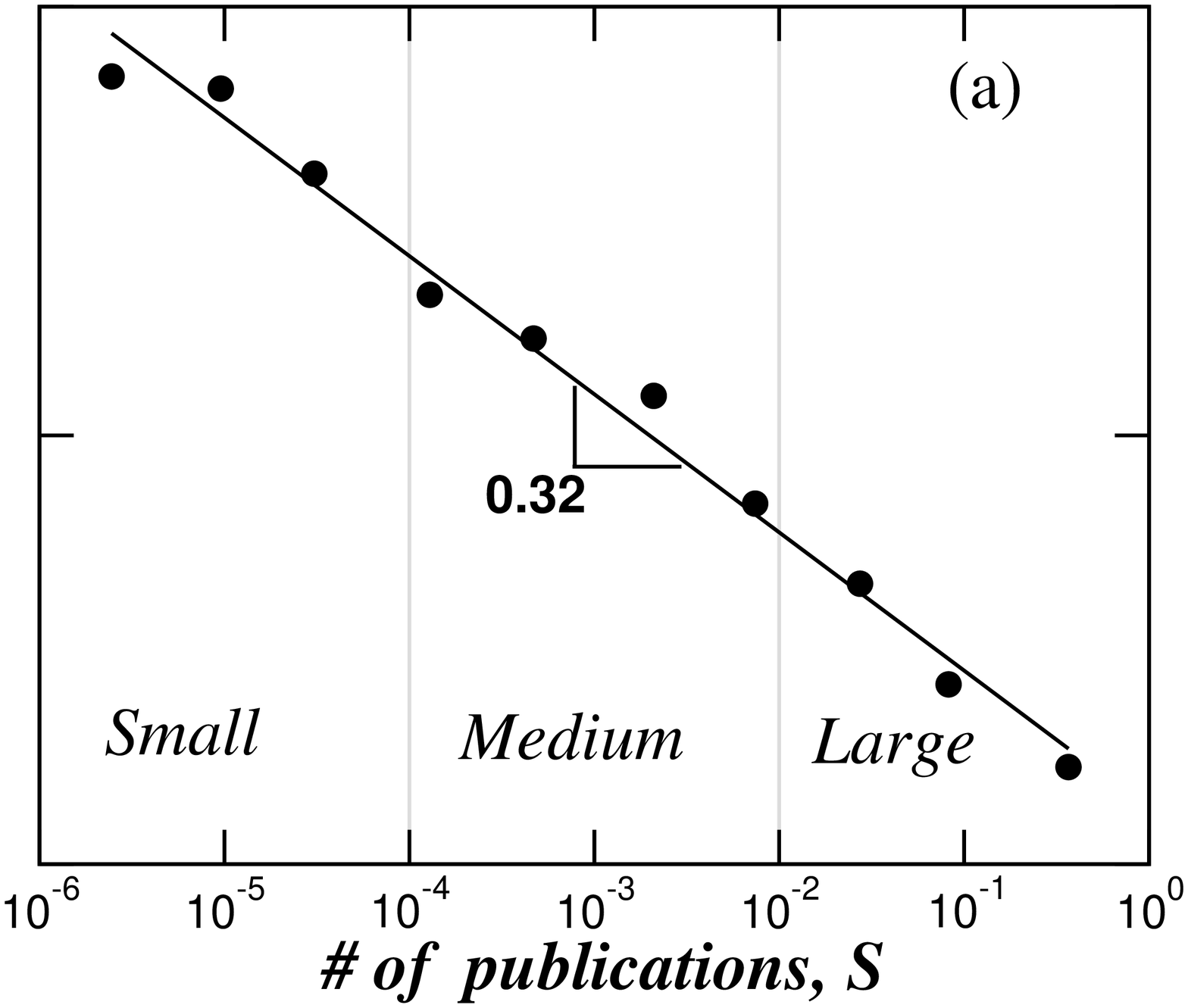}
\end{center}
\begin{center}
\includegraphics[width=0.53\textwidth,angle=-90]{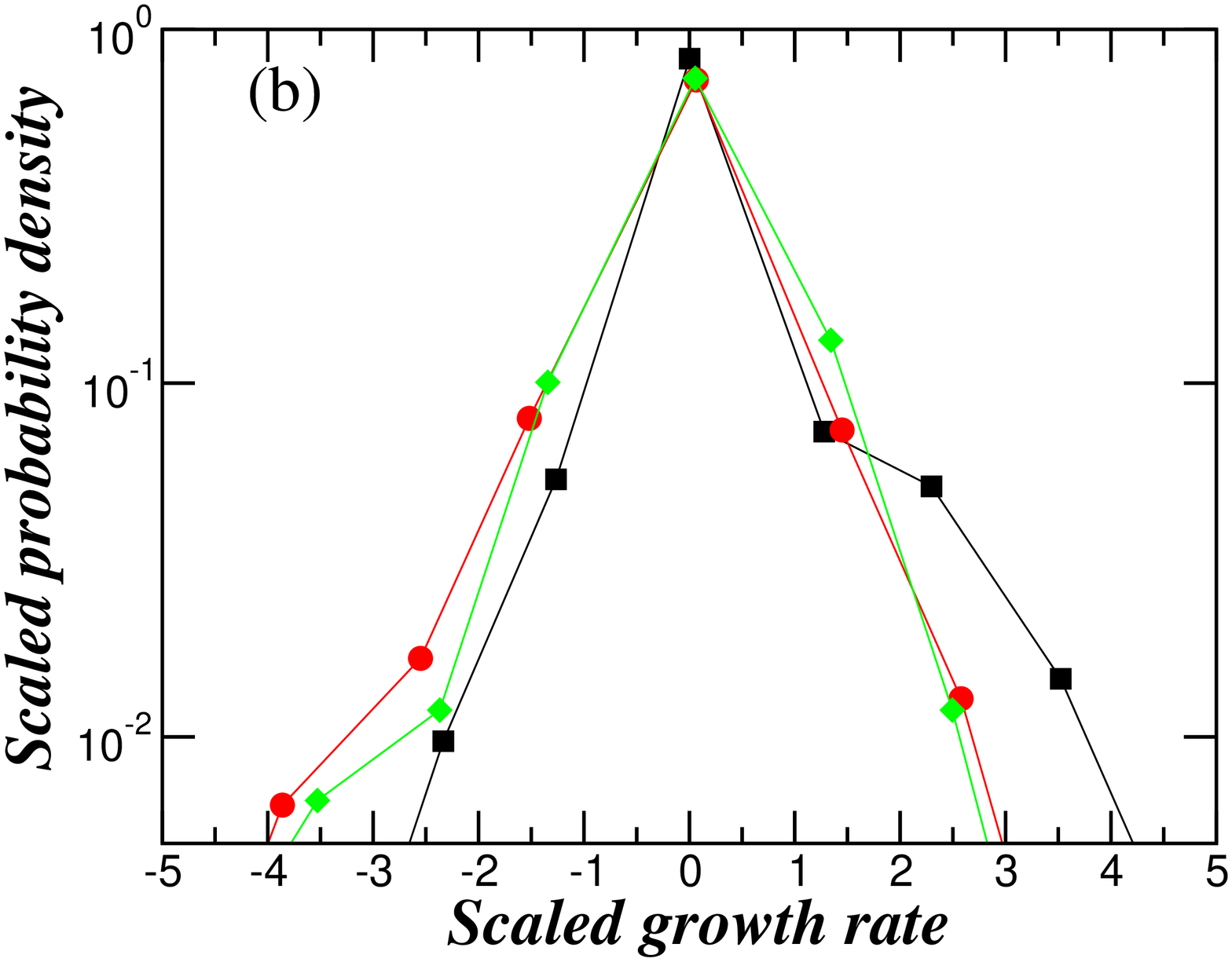}
\end{center}
\vspace*{0.1cm}
\caption{ Fractional counts of world publications. (a) Total world
publication is divided into 10 groups according to size $S$. We find
$\sigma(g|S)$ of the growth rates conditioned on $S$ scales as a power
law i.e., $\sigma(g|S) \sim S^{-\beta} $ with $\beta=0.32$. (b)
Probability distribution of the growth rates of the three sectors scaled
by their standard deviation. Note the collapse of the histograms of the
three sectors.}
\label{Fig3a}
\end{figure}

Figures~\ref{Fig2} and~\ref{Fig3a} present results for the size
distribution of the countries according to the fractional counting
schemes.  Figure~\ref{Fig2} displays the histogram of the logarithm of
the number of publications of 124 countries for the 22 year period
between 1980--2001. We observe that the distribution exhibits a bi-modal
size distribution which implies that the set of 124 countries can be
divided into two classes. In the class with larger sizes we find
countries from the European Union, the North American subcontinent, the
Organization for Economic Co-operation and Development (OECD), and
populous countries such as India, China, and South Africa. In the class
with smaller sizes we find developing countries of the African and South
American continents and countries from the Middle East. The bi-modal
distribution suggests the existence of two different classes of
countries which have an economic and scientific collaboration among
themselves. Note that this result is different from that found for the
GDP of growth of countries (Canning, 1998). In terms of GDP different
countries exhibit a uni-modal distribution, but we see that in terms of
scientific outputs, perhaps because of a more aggressive science policy,
countries exhibit a bi-modal distribution. Analysis applying the two
integer counting schemes generated patterns that are similar to that
obtained with the fractional counting schemes. This feature is also
indicative of the scientific collaboration among different countries in
the two classes observed. One expects that in the case where every
country scientifically collaborates uniformly with every other country
there would not be any segregation into different classes. The
multiplicative growth process in scientific publications is present in
each of these two classes, giving rise to a log-normal distribution,
which is a prediction of Gibrat's theory (Gibrat, 1931) which states
that growth rates of firms are independent and uncorrelated to the firm
size and hence the probability distribution of the firm sizes is
log-normal.

We define the deflated size $S_i(t)$ of the publications of a country $i$
as
\begin{equation}
S_i(t) \equiv \frac{s_i(t)}{\sum_{i=1}^N s_i(t)},
\end{equation}
where $N=124$ and $s_i(t)$ is the number of publications of a country
$i$ in year $t$. The annual growth rate of a country's publication $i$
is defined as
\begin{equation}
g_i(t) = \log S_i(t+\Delta t) - \log S_i(t),
\label{e.growth}
\end{equation}
with $\Delta t = 1$ year. We expect that the statistical properties of
the growth rate $g$ depend on $S$, since it is natural that the
magnitude of the fluctuations $g$ will decrease with $S$.  
We next calculate the standard deviation $\sigma(S)$ of the distribution
of growth rates as a function of $S$.  Figure~\ref{Fig3a}(a) demonstrates
that $\sigma(S)$ decays as a power law
\begin{equation}
\sigma(S) \sim S^{-\beta},
\label{e.sigma2}
\end{equation}
with $\beta = 0.32 \pm 0.05$. To test if the conditional distribution of
growth rates has a functional form independent of the size of the
country, we plot the scaled quantities 
\begin{equation}
p \left( \frac{g}{\sigma(S)} | S \right) ~~~{vs.}
~~~ \frac{g}{\sigma(S)} \,.
\label{e.scaled}
\end{equation}
for 3 different groups partitioned with respect to their size of
publication $S$: small ($S<10^{-4}$), medium ($10^{-4} <S <10^{-2}$),
and large ($S>10^{-2}$). Figure~\ref{Fig3a}(b) shows that the scaled
conditional probability distributions collapse onto a single curve
(Stanley, 1999), suggesting that $p(g | S)$ follows a universal scaling
eq.~\ref{e.scaled}.


FIGURE 1 AND 2 ABOUT HERE


\subsection{Academic Institutions}

\begin{figure}
\narrowtext
\vspace*{0.1cm}
\begin{center}
\includegraphics[width=0.52\textwidth,angle=-90]{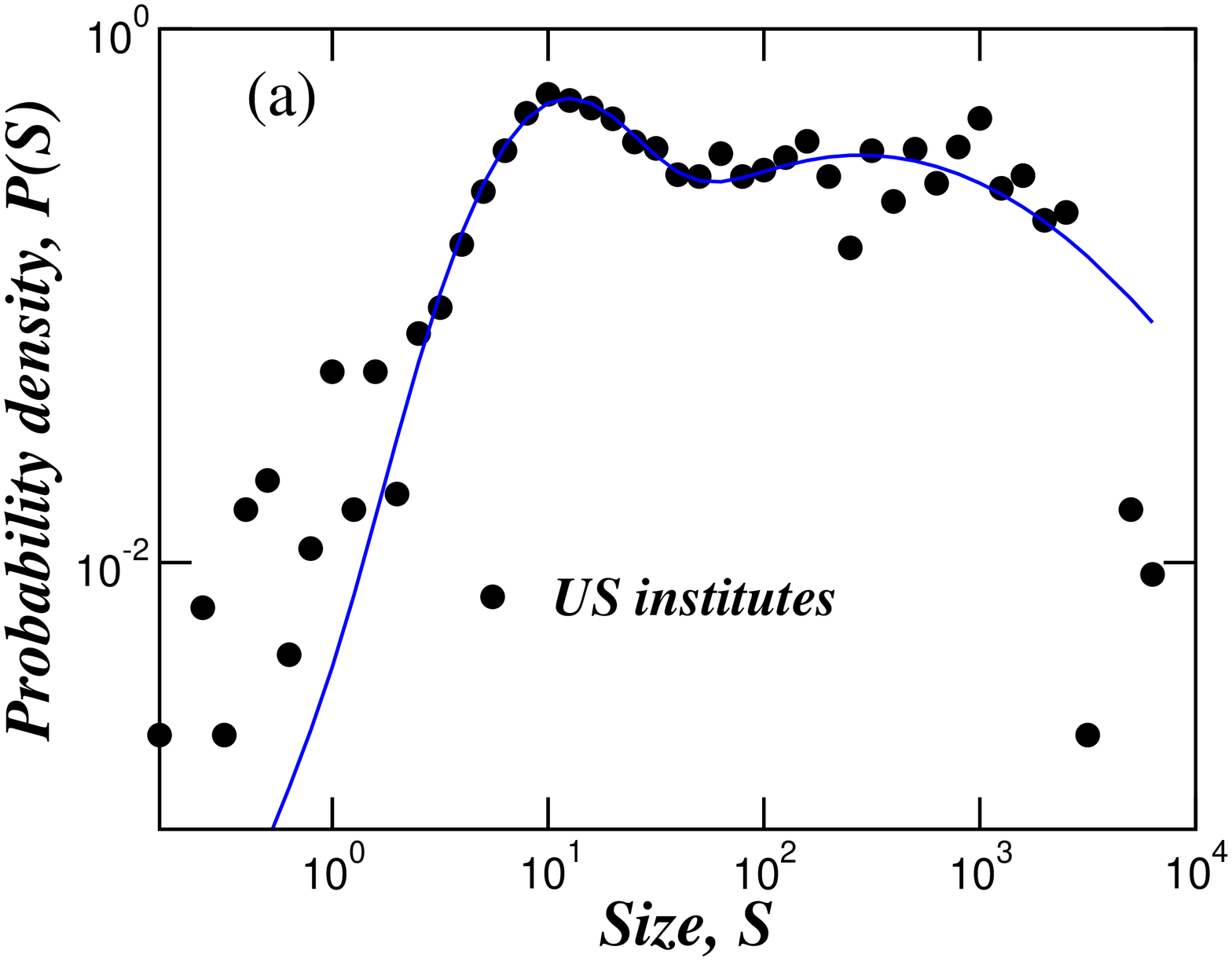}
\end{center}
\begin{center}
\includegraphics[width=0.52\textwidth,angle=-90]{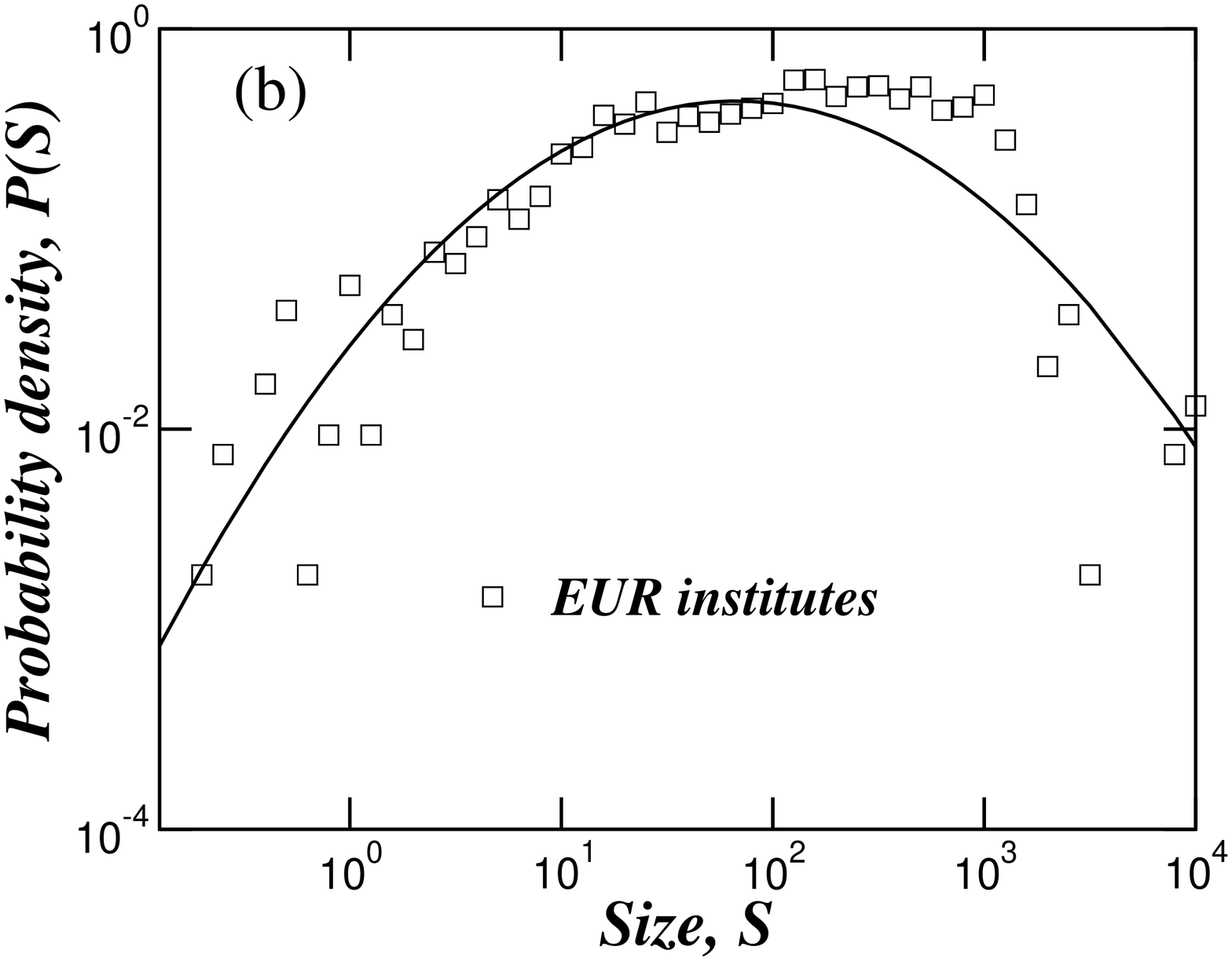}
\end{center}
\vspace*{0.6cm}
\caption{ Histogram of the logarithm of the institutional publication
for (a) 408 USA institutes and (b) 508 EUR institutes measured in the
fractional counting scheme for the 11-year period between 1991--2001.
The full lines are Gaussian fits to the data, which is a prediction of
Gibrat's theory. For EU academic institutions we observe a uni-modal
distribution unlike that observed in distribution of size of publication
for countries. This feature of size distribution is also observed in the
GDP of countries Canning (1998).}
\label{Fig8}
\end{figure}

We now present results for the size distribution of the institutional
publication according to the different regions.  Figure~\ref{Fig8}a
displays the histogram of the logarithm of the number of publications of
408 USA institutes for the 11 year period between 1991--2000. We observe
that the distribution, for EU institutions unlike the US institutions,
exhibits a uni-modal size distribution which was unlike that observed
for publication of countries. Note that this result is similar to that
found for the GDP of growth of countries (Canning, 1998). A possible
conjecture of observing uni-modal distribution as opposed to a bi-modal
distribution of size is a more homogeneous collaboration among
institutes.  The multiplicative growth process in scientific
publications gives rise to a log-normal distribution, which is a
prediction of Gibrat's theory. The distribution for US academic
institutions exhibit a bi-modal rather than a uni-modal pattern. The
values of the scaling parameter $\beta$, however, are statistically
similar in the two academic systems [c. f. Table~\ref{table4}].


FIGURE 3 AND 4 ABOUT HERE


\subsection{Authors}

Next we present results for the size distribution of the Flemish
publication according to the different counting schemes.
Figure~\ref{Fig11} displays the histogram of the logarithm of the number
of publications of 2330 countries for the 21 year period between
1980--2000. We observe that the distribution, for two different counting
schemes, exhibits a uni-modal size distribution which was unlike that
observed for publication of countries. Note that this result is similar
to that found for the GDP of growth of countries (Canning, 1998). In
terms of GDP different countries exhibit a uni-modal distribution, and
we see that in terms of scientific outputs at the level of authors this
feature is similar. This feature is also indicative of the scientific
collaboration among different authors in a uniform way. One expects that
in the case where every author scientifically collaborates uniformly
with every other author there would not be any segregation into
different classes. The multiplicative growth process in scientific
publications gives rise to a log-normal distribution, which is a
prediction of Gibrat's theory. Table~\ref{table4} summarizes the
estimates of scaling exponent $\beta$ [c.f. eq.~\ref{e.sigma}] for
different levels of aggregation. We observe that for different level of
aggregation or for different counting schemes we get statistically
similar values.


TABLE 1 ABOUT HERE


FIGURE 5 AND 6 ABOUT HERE



\section{Deviation from Scaling Laws for Countries}

Next we look at the joint distribution of the relative growth rate and
the relative deviation of $\sigma(S)$ from the scaling laws found in the
previous section. First we define the mean growth rate of a country $j$
as $g_{\mbox{mean}}^j = \frac{1}{21} \sum g_i^j$, where $g_i^j$ is the
growth of country $j$ in year $i=1980,...,2000$. Next we evaluate the
relative growth rate of country $j$ as $g_{\mbox{rel}}^j =
g_{\mbox{mean}}^j/\sigma^j$, where $\sigma^j$ is the standard deviation
of \{$g_{1980}^j,..,g_{2000}^j$\} of country $j$. We then evaluate the
deviation of the countries from the scaling law 
\begin{equation}
\sigma(S) = CS^{-0.37},
\label{e.scale} 
\end{equation}
where $C$ is a constant. We define $\delta\sigma(S_j) = \sigma(S_j) - C
S_j^{-0.37}$, where $S_j$ is the size of country $j$ and then evaluate
$\sigma_{\mbox{rel}}^j \equiv \sigma_{\mbox{rel}}(S_j) =
\delta\sigma(S_j)/\sigma(\delta\sigma(S_j))$, where
$\sigma(\delta\sigma(S_j))$ is the standard deviation of
\{$\delta\sigma(S_1),..,\delta\sigma(S_{124})$\}, evaluated over 124
countries. The scatter plot of
$g_{\mbox{rel}}^j~vs.~\sigma_{\mbox{rel}}^j$ would fall inside a
circular region of 1 standard deviation for countries following the
scaling laws closely. Countries for which
($g_{\mbox{rel}}^j~,~\sigma_{\mbox{rel}}$) falls outside the 2 standard
deviation zone can be hypothesized to pursue a different science and
technology policy than that pursued by the rest of the world with 95\%
probability.

Figures~\ref{Fig4} displays the relative growth rate
$g_{\mbox{\scriptsize rel}}$ plotted against the deviation of $\sigma$
from the best fit line i.e., $\sigma_{\mbox{rel}}$. Circular lines in
the plots mark the different zones of standard deviation in
$\sigma_{\mbox{\scriptsize rel}}$ and $g_{\mbox{\scriptsize
rel}}^j$. Countries falling outside the one standard deviation zone have
deviate significantly from the mean properties of world scientific
outputs. Countries falling in the first quadrant outside the one
standard deviation zone in this plot have positive growth, but the
standard deviation in the growth rate implies that the fluctuation in
the growth is high. Countries falling in the second quadrant have high
positive growth and also less standard deviation in growth, indicating a
more stable growth process. Countries falling outside the one standard
deviation zone in this quadrant are quickly developing
countries. Scientific research from these countries may produce newer
fields resulting in high positive growth and bigger
fluctuations. Countries outside the one standard deviation zone in the
third quadrant are countries with strongly decaying science
policies. Both the standard deviation of growth and the growth is
negative, suggesting a very strong decay. Countries in the fourth
quadrant outside the one standard deviation zone have higher standard
deviation in growth, but the growth itself is negative. The countries in
this quadrant have a chance to move over to the first or second quadrant
because of higher fluctuations. These are the newly developed countries
which may be recently investing in scientific research.


FIGURE 7 ABOUT HERE


Figures~\ref{Fig5} display the standard deviation $\sigma$ of the growth
rates of all 124 countries plotted as a function of $S$, in two periods
between 1981-1990 and 1991-2000 for (a) fractional, (b) integer type I,
(c) integer type II counting schemes. Comparison of scaling laws in
these two consecutive decades may be indicative of any policy or
political regime changes that countries possibly have undergone. We
observe that the countries have identical scaling laws in the two
consecutive decade.

Next we study the deviation of $\sigma(S)$ from the best fit line in for
the two 11 year periods between (a) 1980-1990 and (b) 1991-2000
(c.f. Fig.~\protect\ref{Fig4}, which is the entire 22 year
period). We observe that China and South Korea had a very high deviation
of growth rate from the average growth rate of world publication during
the period 1980-1990. During the second half of the analysis period we
observe both countries as deviating less from the average world
publication grow rate. We also observe the growth rate of USA as
becoming more stable and moving inside the 1 standard deviation zone in
the 2nd half of the analysis period. Dramatic policy changes are also
observed for countries such as Iran which shift from the negative 2
standard deviation zone to the positive 2 standard deviation zone during
these two decades. Developing countries such as India become more stable
in terms of their science policy and move inside the 1 standard
deviation zone and countries such as Japan become more deviant and more
within to the 1 standard deviation zone.


FIGURE 8 ABOUT HERE


\section{Discussion }

We have described a research approach that may be quite new in the field
of scientific policy and that may shed light on the behavior and
characteristics of S\&T systems. Understanding these processes and the
data characterizing them is of great relevance not only for S\&T studies
but also for science policy. Indeed, countries are increasingly
stressing performance because research funding is becoming more and more
an instrument for safeguarding long term economic
competitiveness. Scientific research can be modeled as an input-output
process, according to which inputs such as the stocks of scientific
knowledge and existing techniques, skilled personnel, scientific
instruments, recruited personnel, and financial resources, are
transformed by conceptual, experimental, and technical work of
scientists into outputs, particularly scientific contributions, to a
discipline in the form of new scientific knowledge, techniques, and
trained scientists.

Our study deals with scientific performance or scientific
excellence. National governments, particularly in OECD countries, make
large investments in basic scientific research. During the past decades,
the need for accountability in scientific research and research student
training has increased strongly. As indicated earlier and observed
empirically, this type of aggressive science policy by a group of
countries may be a cause of the bi-modal distribution of sizes.

Our studies on the EU and the institutions reveal another special
characteristic observed within the EU but not in US institutions. The
uni-modal size distribution is indicative of a homogeneous collaboration
among institutes of all size. A bi-modal distribution which is observed
in US institutions is indicative of a clustering effect of institutes of
two different size classes. Whether or not we observe this clustering
effect in collaboration among institutes in EU and USA the scaling
parameter of growth remains statistically similar to that observed for
countries. It is indeed remarkable that for all levels of aggregation
i.e., from countries to research institutes to authors, the scaling
parameter of growth as a function of size remains statistically
comparable. These important results observed in the scientific output of
countries and research institutes were not observed in the GDP of
countries or other S\&T input output indicators like citation.

In our macroscopic analysis in which we study the statistical properties
of the growth rates in the annual number of articles published by a
country, a certain statistical regularity was found between a country's
standard deviation and its total volume of published articles.  The
standard deviation as a function of the total number of articles
published decays as a power law. The exponent in the power law equation
is denoted in statistical physics as the scaling exponent.  A closer
inspection of the results reveals that for some countries, the standard
deviations in their annual growth rates deviate substantially from the
expected scores given by the total number of papers they published. The
significance of such a deviation and what it can teach us about the
efficiency of the various national research systems will be addressed in
the next phase of our research.

We thank X.~Gabaix, S.~Havlin, M.~Salinger, for helpful discussions and
suggestions, and NSF for financial support.

\newpage

\eject

\begin{table}
\caption{Scaling Exponent for Different Levels of Aggregation}
\begin{tabular}{|c|c|c|}
\hline
\hline
Level of    & Counting Schemes & $\beta$ \\
Aggregation &                  &         \\
\hline
Countries         &                  &                 \\ 
                  & Fractional Count & 0.32 $\pm$ 0.05 \\
                  & Integer Count I  & 0.32 $\pm$ 0.05 \\
                  & Integer Count I  & 0.34 $\pm$ 0.05 \\
\hline
Institutes        &                  &                 \\
EU                & Fractional Count & 0.39 $\pm$ 0.05 \\
USA               & Fractional Count & 0.30 $\pm$ 0.05 \\
EU + USA Combined & Fractional Count & 0.35 $\pm$ 0.05 \\
\hline
Flemish Authors   &                  &                 \\ 
                  & Fractional Count & 0.28 $\pm$ 0.05 \\
                  & Integer Count    & 0.22 $\pm$ 0.05 \\
\hline
\hline
\end{tabular}
\vspace{0.2cm}
\label{table4}
\end{table}

\eject
\newpage


\newpage

\newpage

\newpage

\begin{figure}
\narrowtext
\vspace*{0.6cm}
\begin{center}
\includegraphics[width=0.60\textwidth,angle=-90]{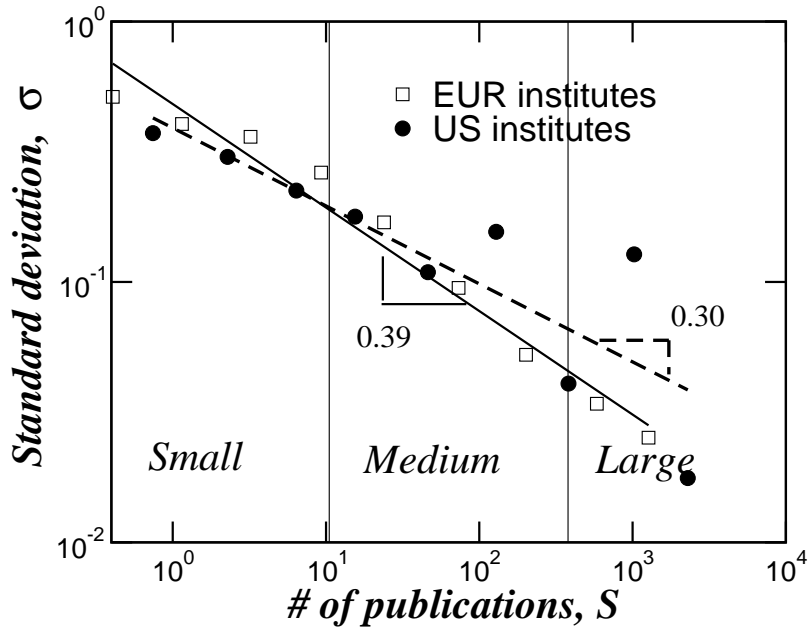}
\end{center}
\vspace*{0.1cm}
\caption{ Total EU publication (square) is divided into 10 groups
according to size $S$. We find $\sigma(g|S)$ of the growth rates
conditioned on $S$ scales as a power law i.e., $\sigma(g|S) \sim
S^{-\beta} $ with $\beta=0.39$. Total US publication (circle) is divided
into 10 groups according to size $S$. We find $\sigma(g|S)$ of the
growth rates conditioned on $S$ scales as a power law i.e., $\sigma(g|S)
\sim S^{-\beta} $ with $\beta=0.30$.}
\label{Fig9a}
\end{figure}

\newpage

\begin{figure}
\narrowtext
\vspace*{0.1cm}
\begin{center}
\includegraphics[width=0.52\textwidth,angle=-90]{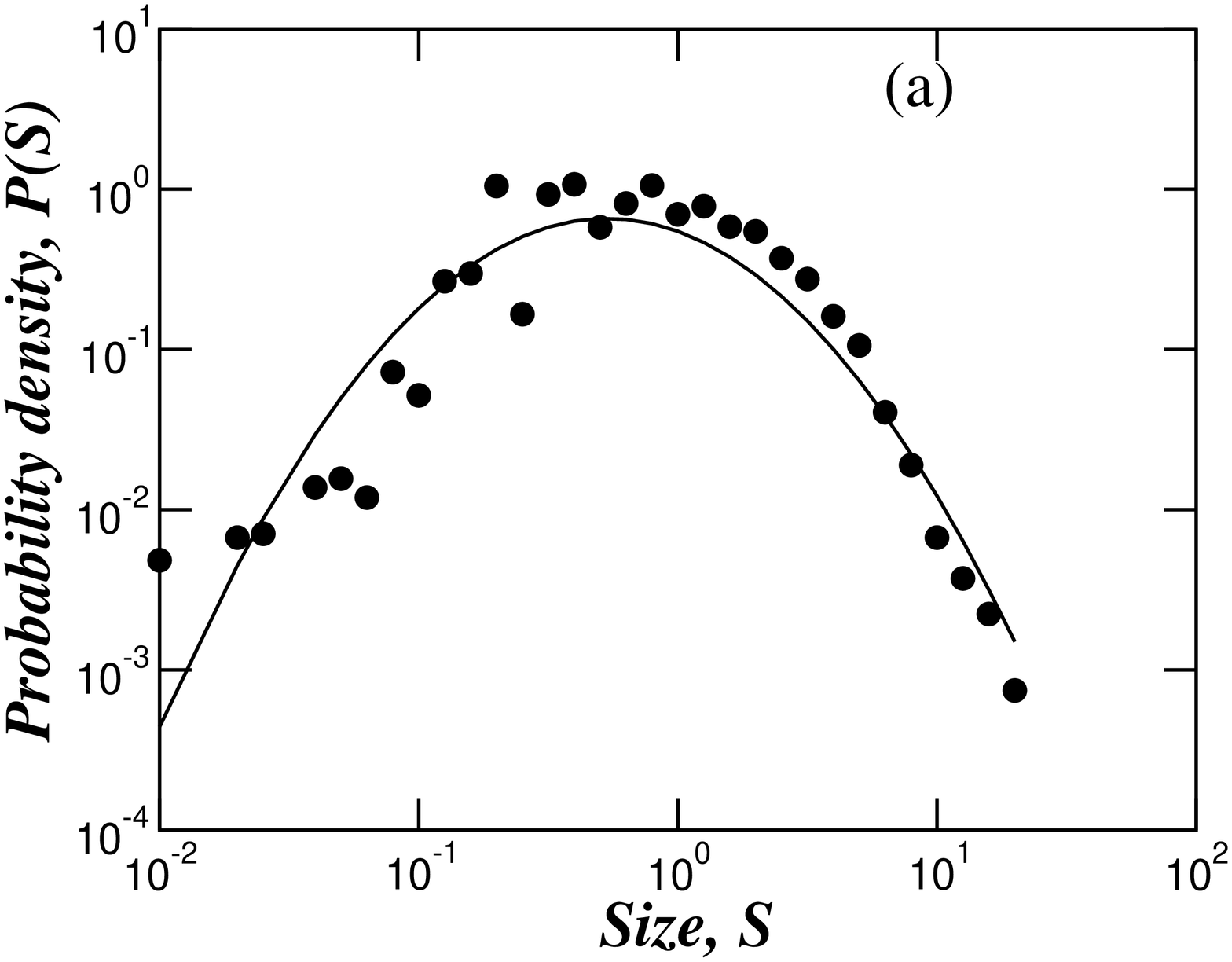}
\end{center}
\begin{center}
\includegraphics[width=0.52\textwidth,angle=-90]{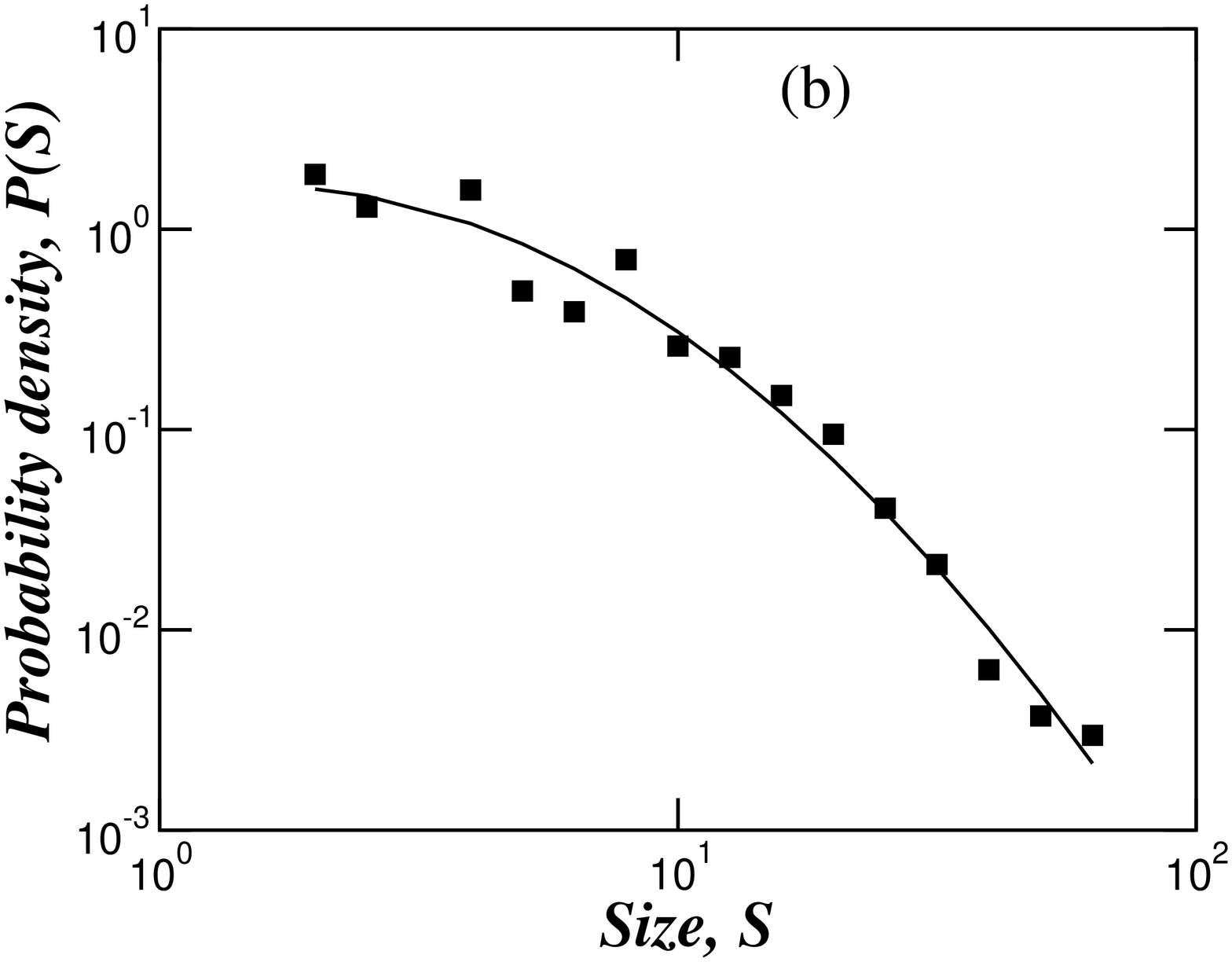}
\end{center}
\vspace*{0.6cm}
\caption{ Histogram of the logarithm of the (a) fractional count, (b)
integer count of number of publications of 2330 Flemish authors for the
21-year period between 1980--2001.  The full lines are Gaussian fits to
the data, which is a prediction of Gibrat's theory which states
that growth rates of firms are independent and uncorrelated to the firm
size and hence the probability distribution of the firm sizes is
log-normal.}
\label{Fig11}
\end{figure}

\newpage

\begin{figure}
\narrowtext
\vspace*{0.6cm}
\begin{center}
\includegraphics[width=0.52\textwidth,angle=-90]{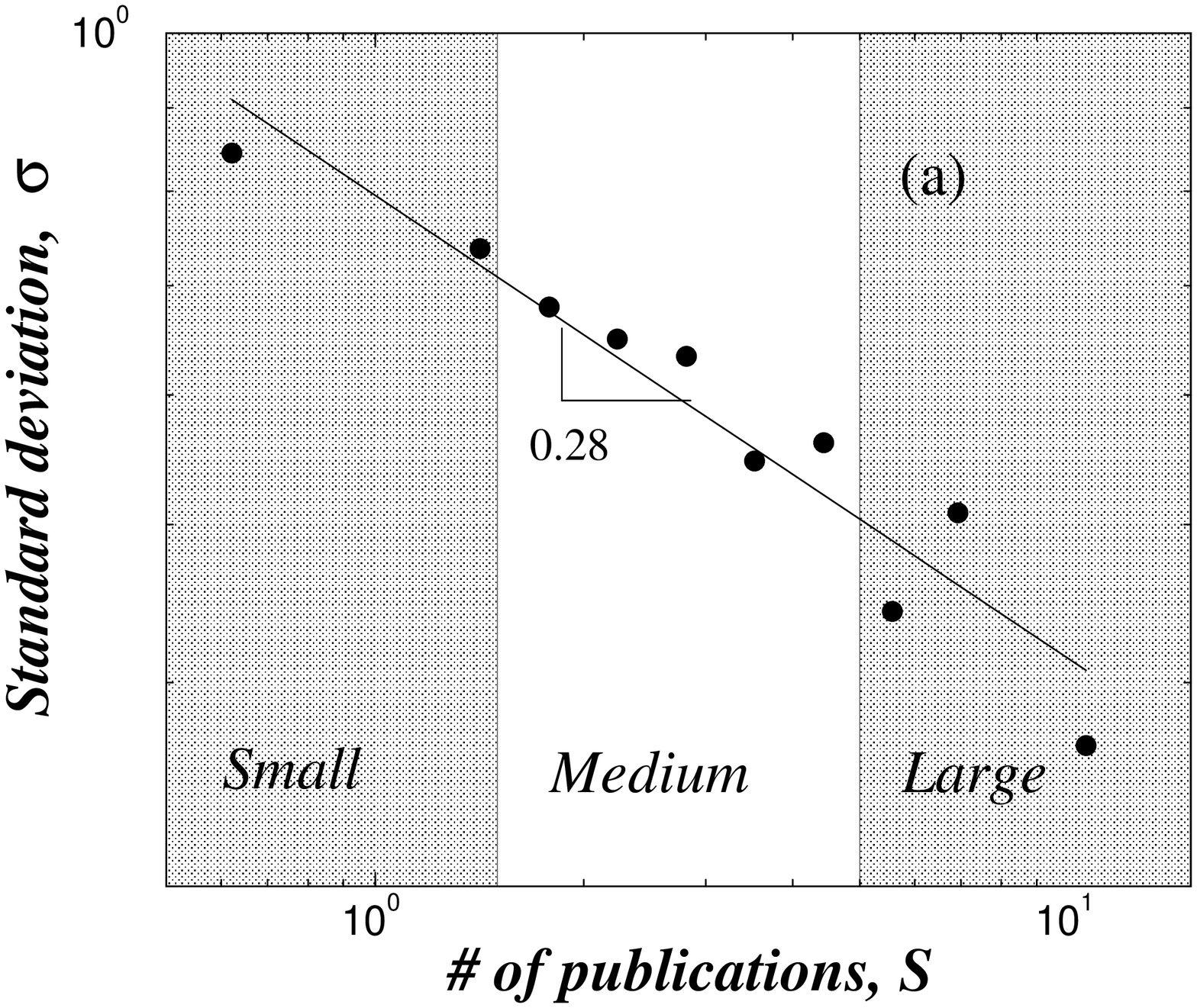}
\end{center}
\begin{center}
\includegraphics[width=0.52\textwidth,angle=-90]{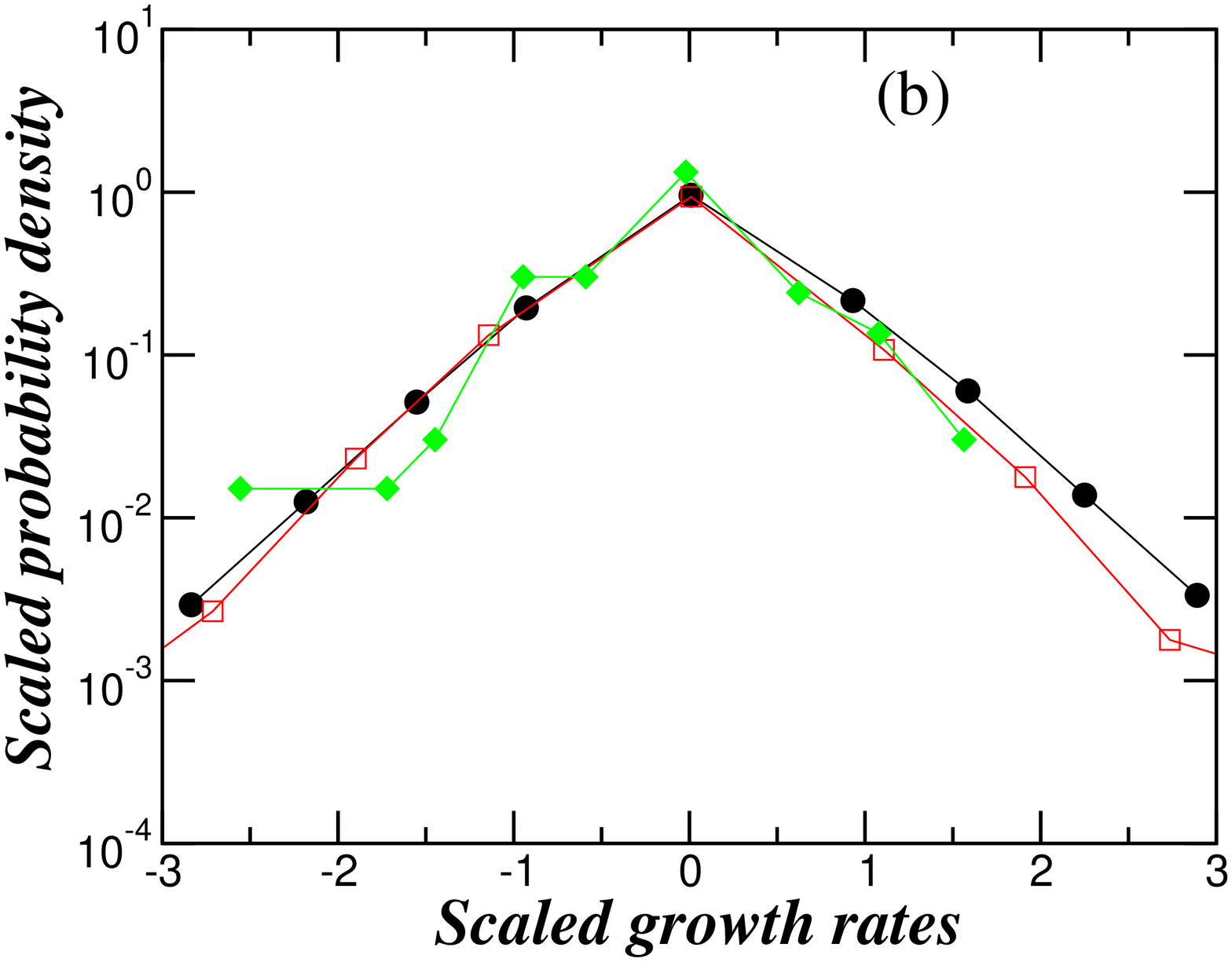}
\end{center}
\vspace*{0.1cm}
\caption{ Fractional counts of Flemish publications. (a) Total Flemish
publication is divided into 10 groups according to size $S$. We find
$\sigma(g|S)$ of the growth rates conditioned on $S$ scales as a power
law i.e., $\sigma(g|S) \sim S^{-\beta} $ with $\beta=0.28$. (b)
Probability distribution of the growth rates of the three sectors scaled
by their standard deviation. Note the collapse of the histograms of the
three sectors.}
\label{Fig12a}
\end{figure}

\newpage

\begin{figure}
\narrowtext
\vspace*{0.9cm}
\begin{center}
\includegraphics[height=0.60\textwidth,angle=-90]{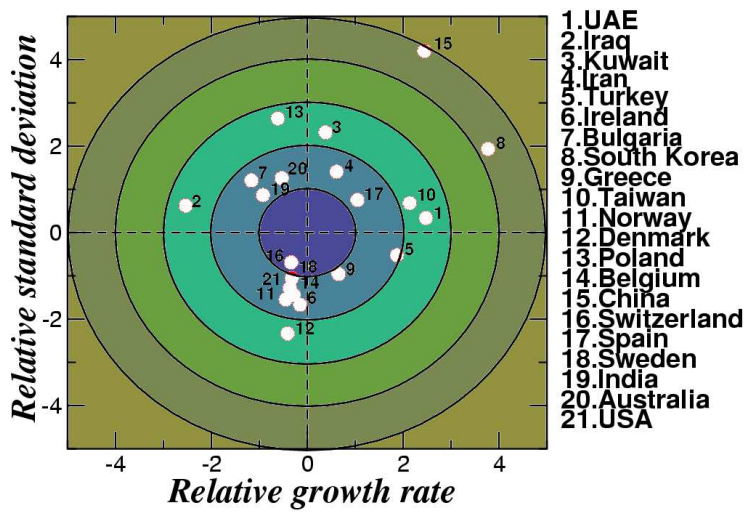}
\end{center}
\vspace*{0.1cm}
\caption{ Scaled growth rates versus the scaled deviation of $\sigma(S)$
from the best fit line for the first few countries ranked (based on the
total annual publication size) within 30. Observe that countries outside
the 2 $\sigma$ contour deviate from the $\sigma~vs.~ S$ scaling law with
$>95\%$ confidence. Note that developing countries such as South Korea
and China have a very high positive growth rate.}
\label{Fig4}
\end{figure}

\newpage

\begin{figure}
\narrowtext
\vspace*{0.6cm}
\begin{center}
\includegraphics[width=0.30\textwidth,angle=-90]{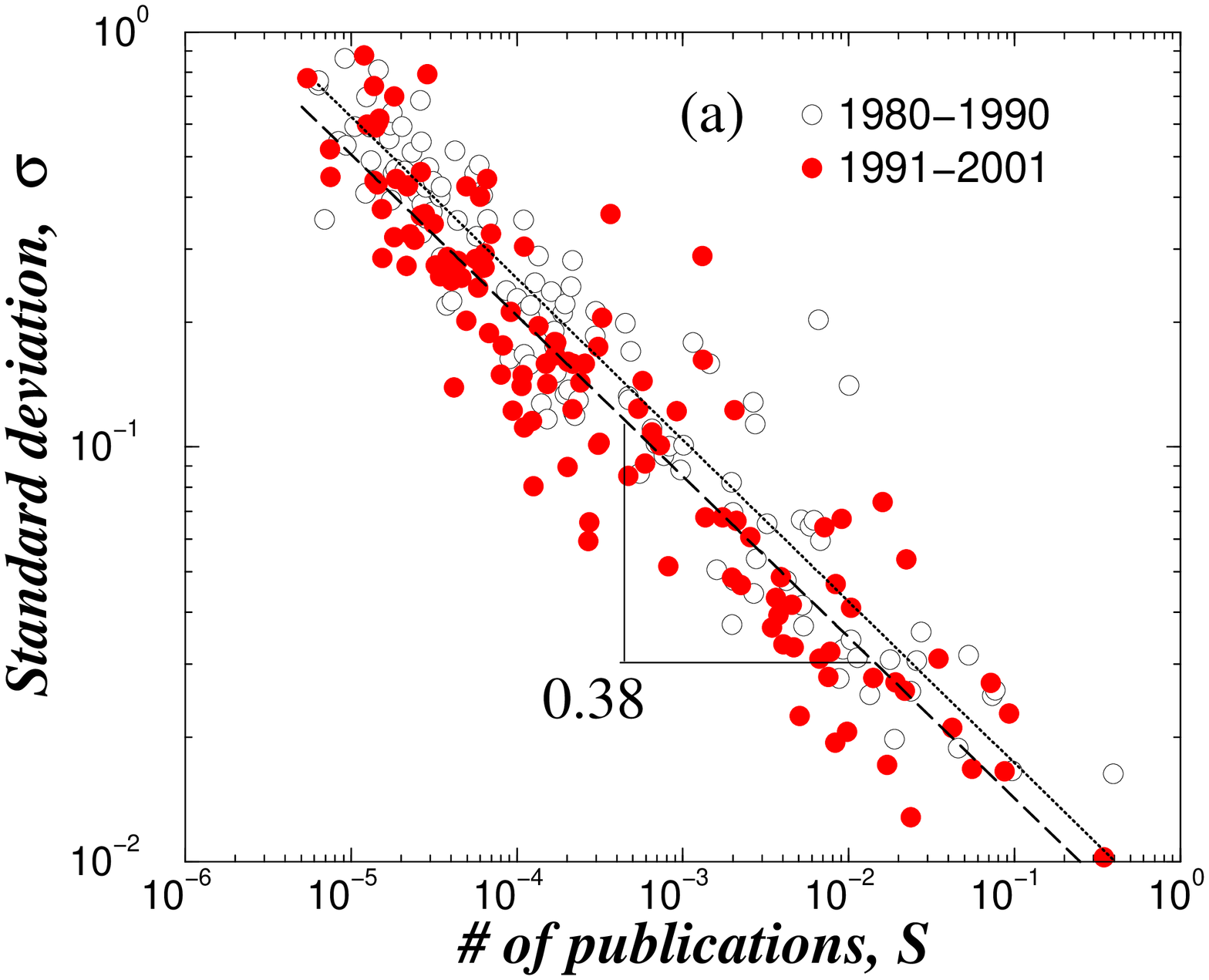}
\end{center}
\begin{center}
\includegraphics[width=0.30\textwidth,angle=-90]{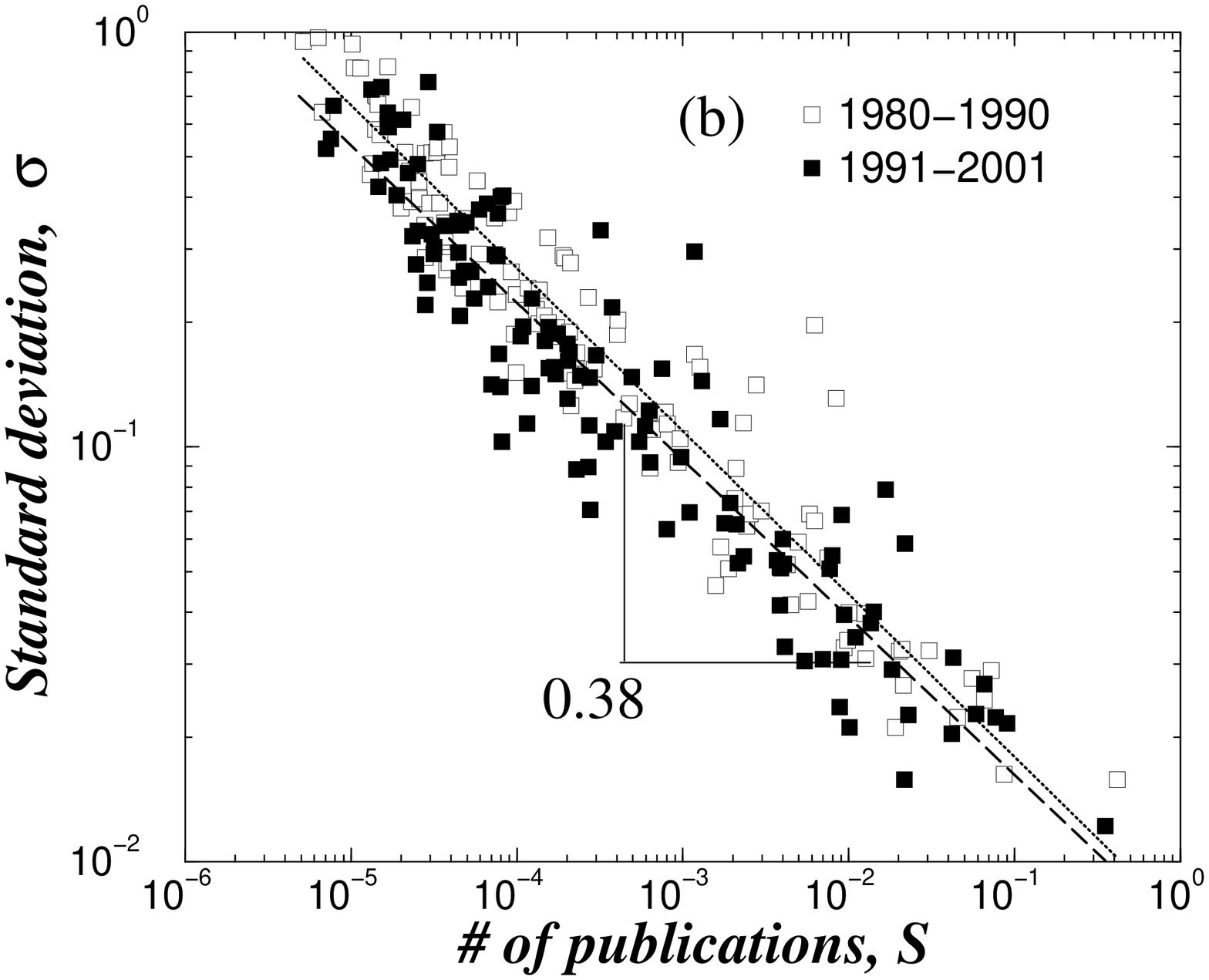}
\end{center}
\begin{center}
\includegraphics[width=0.30\textwidth,angle=-90]{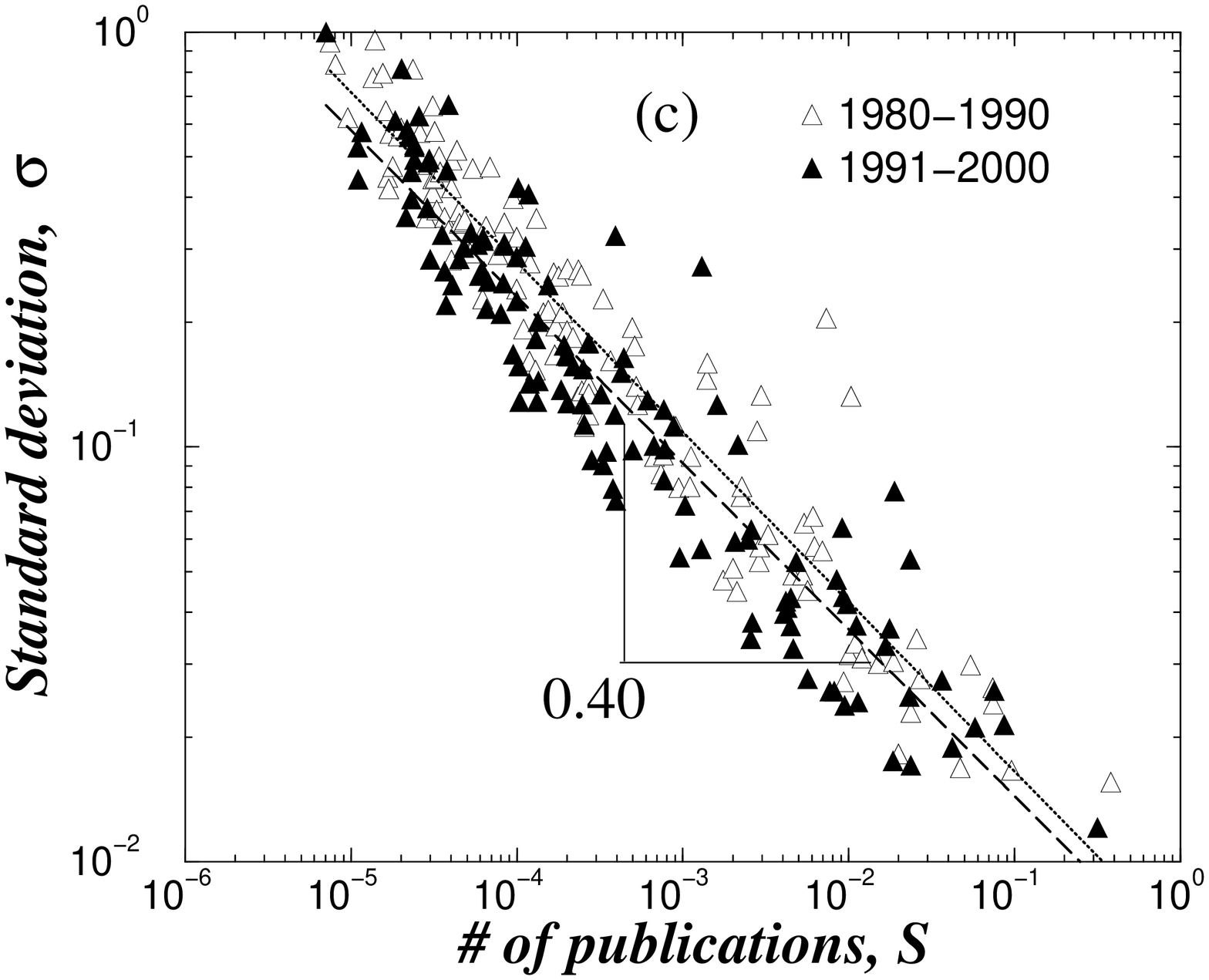}
\end{center}
\vspace*{0.1cm}
\caption{Standard deviation $\sigma$ of the growth rates of all 124
countries plotted as a function of $S$, in periods between 1981-1990 and
1991-2000 for (a) fractional, (b) integer type I, (c) integer type II
counting schemes. Comparison of scaling laws in these two consecutive
decades may be indicative of any policy or political regime changes
countries might have undergone. The deviation from scaling for the
different counting schemes are indicative of changes in institutional or
international collaborations.}
\label{Fig5}
\end{figure}

\newpage




\end{document}